\documentstyle[11pt,newpasp,twoside,epsf]{article}
\begin{document}

\title{Testing Cosmic Acceleration with Type Ia Supernovae}
\author{Saurabh Jha}
\affil{Harvard-Smithsonian Center for Astrophysics, 60 Garden Street,
Cambridge, MA 02138, USA}
\author{and the High-Z Supernova Search Team}
\affil{PI: Brian P. Schmidt, Mt. Stromlo
Observatory, Cotter Road, Weston Creek, ACT 2611, Australia, \\ 
{\rm http://cfa-www.harvard.edu/cfa/oir/Research/supernova/HighZ.html}}
\begin{abstract}
We discuss recent evidence for an accelerating Universe from
measurements of type Ia supernovae at high redshift, and describe
tests of various systematic effects such as extinction and evolution
that could be biasing the cosmological result. Continued observations
of these objects, both over a wider wavelength region and at higher
redshift, should provide strong evidence in favor or against the
accelerating Universe hypothesis.
\end{abstract}

\section{Introduction}

Recent observations of type Ia supernovae (SN Ia) at high redshift ($z
\ga 0.3$) by two groups provide evidence that the Universe is
accelerating at the current epoch (Riess et al.~1998, hereafter R98;
Perlmutter et al.~1999, hereafter P99). This result implies the
existence of a constituent of the Universe, either the cosmological
constant, $\Lambda$, or something else with similarly negative
pressure capable of accelerating the expansion, generically dubbed
``dark energy'' (see Riess 2000 for a review).

Restricting ourselves to a cosmological-constant model, the current
constraints on cosmological parameters from SN Ia and cosmic microwave
background (CMB) observations are shown in Figure 1. The shaded SN Ia
region corresponds to the 99.7\% confidence region on the matter
density and vacuum energy density ($\Omega_M$, $\Omega_\Lambda$) from
a combination of both groups' published data (R98; P99), ensuring the
sets of supernovae used were independent. Similarly, the shaded CMB
region corresponds to the same 99.7\% confidence region on these
parameters from the latest CMB results, including the published
BOOMERANG-98 and MAXIMA-1 data sets (Jaffe et al.~2000). The combined
confidence region is shown with the contours, representing 68.3, 95.4
and 99.7\% confidence levels. The combined constraints rule out a
flat, matter-dominated Universe ($\Omega_M = 1$, $\Omega_\Lambda =
0$), as well as an open Universe with no cosmological constant (e.g.,
$\Omega_M = 0.3$, $\Omega_\Lambda = 0$), at high statistical
significance. In this framework, the data clearly favor a significant
and even dominant fraction of the energy density of the Universe in
the cosmological constant. Even allowing for more exotic possibilities
such as quintessence, the SN Ia data still require a dark energy
component to cause acceleration of the Universe (Garnavich et
al.~1998; P99).

\begin{figure}
\plotone{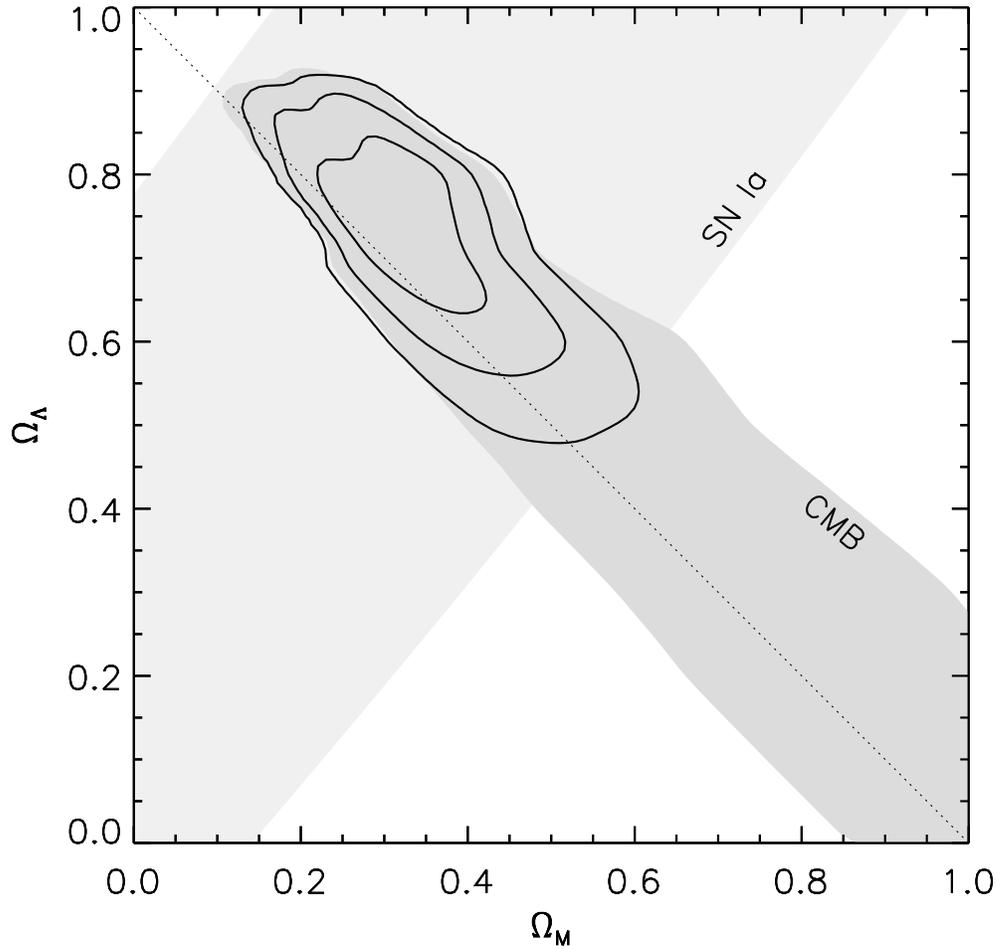}
\caption{Cosmological constraints from SN Ia and CMB data. The SN Ia
shaded region marks the 99.7\% confidence region from an independent
combination of SN Ia data from the High-Z SN Search Team (R98) and the
Supernova Cosmology Project (P99). The shaded CMB region marks the
same confidence region from the latest CMB results (Jaffe et
al.~2000). The combination of the two constraints results in the 68.3,
95.4 and 99.7\% solid contours shown.}
\end{figure}

How secure is this result? The SN Ia data support an accelerating
Universe at a high level of statistical confidence, but as is always
the case, systematic uncertainties then become the primary
concern. The observational result is simple: SN Ia at $z \simeq 0.5$
appear approximately 50\% fainter than expected for a flat,
matter-dominated Universe, and about 25\% fainter than expected for an
open Universe with $\Omega_M = 0.3$. Both groups observing
high-redshift SN have explored a number of systematic effects that
could bias the measurements (R98; P99), including Malmquist bias,
sample contamination and gravitational lensing. None of these
particular effects reconcile the data with a decelerating (or even
coasting) Universe.

There are two other potential sources of systematic uncertainty that
are of obvious concern. First, a natural candidate to explain the
observed faintness of the high-redshift supernovae is extinction by
interstellar dust. Second, evolution of the intrinsic properties of SN
Ia, or a change in the population of SN Ia observed at high-redshift
relative to those nearby, could bias the results. In this paper we
describe methods of testing these possibilities from measurements of
SN Ia themselves. We are in the midst of carrying out these tests; the
results will either bolster confidence in our current cosmological
paradigm or present a serious challenge to the picture, by teaching us
something new about cosmic dust or supernova evolution.

\section{Extinction}

Nearby SN Ia clearly show effects of dust extinction from the Milky
Way and the SN host galaxy (Hamuy et al.~1996; Riess, Press, \&
Kirshner 1996; Phillips et al.~1999; Jha et al.~1999). Extinction by
normal dust grains reveals itself as a color excess, preferentially
extinguishing bluer light. Both groups observing high-redshift SN Ia
now measure light curves in at least two passbands, typically
rest-frame $B$ and $V$, allowing for a measurement of any color excess
and a correction for extinction using a standard reddening law with
$R_V = 3.1$ (Cardelli, Clayton, \& Mathis 1989) performed either on an
individual supernova by supernova basis or in the mean.

The results from both groups to date suggest that extinction by normal
dust is not the reason for the apparent faintness of high-redshift SN
Ia. The color excess $E(B-V)$ for the high-redshift objects is
consistent with zero in both the R98 and P99 samples\footnote{Falco et
al.~(1999) point out that the mean color excess of the R98 sample is
\emph{negative} at the 1 to 2$\sigma$ level of significance, meaning
the SN Ia at high redshift may be \emph{bluer} than the nearby
sample. The more precise observations described here will confirm or
modify this result.}. Furthermore, systematic uncertainties are
unlikely to change this result significantly, the ``worst-case''
plausible color excess is likely no more than 0.03 or 0.04 mag, which
for normal dust grains ($R_V = 3.1$) implies an extinction of at most
$\sim 0.1$ mag, not enough to disfavor an accelerating Universe.

What about sources of opacity along the line of sight other than
normal dust grains? For instance, dust which reddens less than normal
(i.e., a distribution of dust grains with high $R_V$) could extinguish
enough light to reconcile the SN Ia data with a non-accelerating
Universe without measurable reddening. Nonetheless, the current SN
data can still constrain such ``grey'' dust. The scatter of the SN Ia
measurements at high redshift does not show any increase relative to
the nearby sample beyond that caused by measurement uncertainties
(R98; P99). This rules out the existence of grey dust that is patchy
like normal dust; if some of the SN were extinguished by grey dust
more or less than the others, the scatter at high redshift would
increase. Thus, any grey dust dimming the light of high redshift SN
must be relatively uniform, similarly extinguishing the light from
\emph{every} SN. Even with such strong constraints, there are still
models for dust which could behave this way; Aguirre (1999) and
Aguirre \& Haiman (2000) have proposed that \emph{intergalactic} grey
dust could exist in sufficient quantities to dim the light from
distant SN Ia by 0.25 mag, without violating other astrophysical
constraints.

How do we detect such grey dust? Fortunately, any realistic model of
such grains leads to dust that is not completely grey; by observing
distant SN Ia over a wide wavelength range we can detect reddening by
such ``grey'' dust. Riess et al.~(2000) showed that observations of a
high-redshift SN in the rest-frame $B$, $V$, and $I$ bands could begin
to detect the effects of grey dust. However, the data in that paper
were not sufficient to provide a strong constraint, and it was clear
that additional, precise measurements were required. We in the High-Z
SN Search Team are in the process of making such measurements, by
extensively observing a sample of seven SN Ia at $z \simeq 0.5$ from
the ground and with \emph{HST}, covering rest-frame $UBVRI$ (which
approximately corresponds to observer-frame $VRIZJ$).

Figure 2 shows how these observations constrain extinction. The
curves show the expected color excess as a function of wavelength
(i.e., $E(U-B)$, $E(U-V)$, etc.) for different extinction models,
normalized to produce an extinction $A_V = 0.25$ mag necessary to
explain the R98 and P99 data in a non-accelerating Universe. The
normal dust curves correspond to dust grains with $R_V = 3.1$, while
the grey dust curves correspond to two models presented by Aguirre
(1999). We also show the expectation of zero color excess if there
were no dust present. The error bars in the figure indicate the
precision with which we can determine the color excesses for each
supernova based on the details of our observing strategy and
measurement errors. With seven SN Ia these observations should
definitively show whether the apparent faintness of high-redshift
SN is due to extinction by dust, grey or otherwise.

\begin{figure}
\plotfiddle{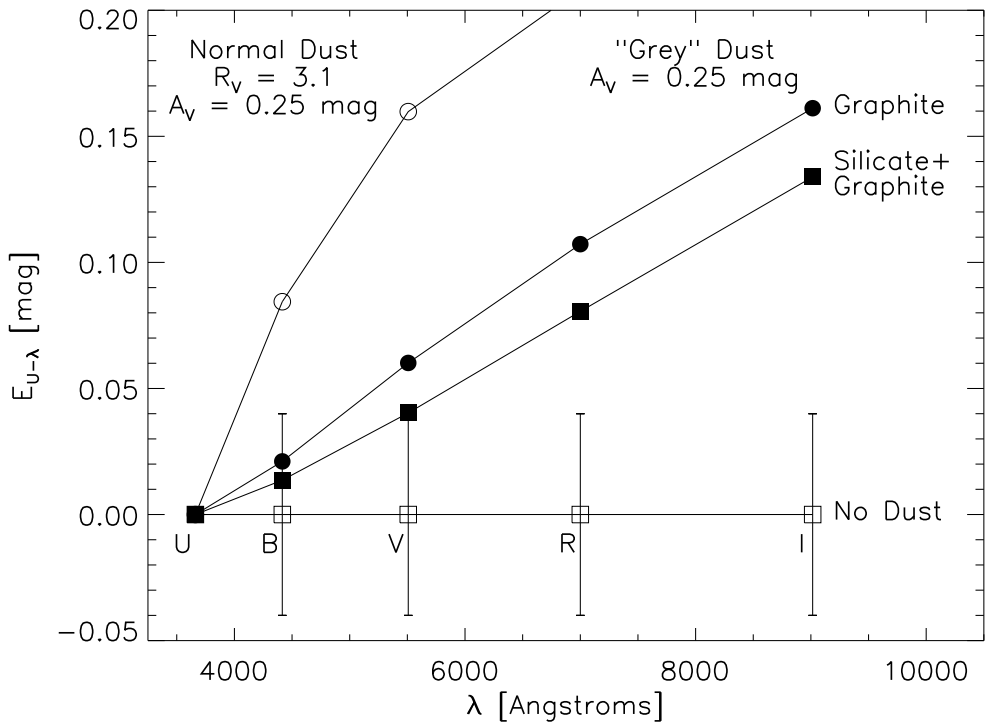}{3.5in}{90}{55}{55}{220}{0}
\caption{Color excess curves for various dust models, including normal
and ``grey'' dust. The curves are normalized to produce $A_V =
0.25$~mag, necessary to reconcile the SN Ia results with a
non-accelerating Universe. Also shown are the uncertainties expected
from measurements of one SN Ia at $z \simeq 0.5$ from rest-frame
$UBVRI$ ground-based and \emph{HST} observations currently
underway. See text for details.}
\end{figure}

Our campaign to discover and followup these $z \simeq 0.5$ SN Ia is
well underway. We discovered a large number of candidates during our
searches at the Canada-France-Hawaii 3.6m telescope and at the Cerro
Tololo Inter-American Observatory 4m telescope, with template runs in
late September 2000 and search runs in late October 2000 (Schmidt et
al.~2000). With subsequent spectroscopy of these candidates from the
European Southern Observatory 8.2m ANTU Very Large Telescope, we were
able to choose seven SN Ia for the intensive followup with \emph{HST}
and ground-based telescopes. We list these supernovae in Table
1. Figure 3 shows one of these objects as imaged by \emph{HST}.

\begin{table}
\caption{SN Ia currently being studied by the High-Z SN Search Team}
\vspace{0.1in}
\begin{tabular}{lcccc}
\tableline
SN Ia & R.A. (2000) & Dec. (2000) & Discovery $R$ Mag & $z$ \\
\tableline
SN 2000dy (Elmo) & 23:25:35.93 & $-$00:22:34.0 & 22.7 & 0.61 \\
SN 2000dz (PlasticMan) & 23:30:41.36 & $+$00:18:42.7 & 23.1 & 0.50 \\
SN 2000ea (RubberDucky) & 02:09:54.02 & $-$05:28:17.8 & 23.3 & 0.42 \\
SN 2000ec (Submariner) & 02:11:32.03 & $-$04:13:56.1 & 22.7 & 0.47 \\
SN 2000ee (InvisibleWoman) & 02:27:34.53 & $+$01:11:49.4 & 22.6 & 0.47 \\
SN 2000eg (WonderWoman) & 02:30:21.05 & $+$01:03:48.5 & 22.5 & 0.54 \\
SN 2000eh (Penguin) & 04:15:02.44 & $+$04:23:18.1 & 22.4 & 0.49 \\
\tableline
\tableline
\end{tabular}
\end{table}

\begin{figure}
\plotfiddle{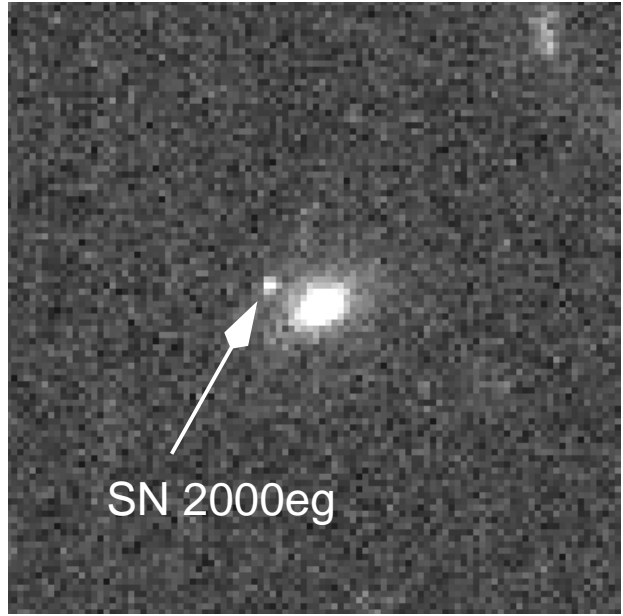}{3in}{0}{80}{80}{-100}{0}
\caption{November 2000 \emph{HST} $R$-band image of SN 2000eg at
$z~=~0.54$. The field is 10 arcseconds on a side.}
\end{figure}

In addition to settling the question of extinction, the full $UBVRI$
light curves for these high-redshift SN Ia will allow us to address
other questions. For instance, precise measurements of the SN color
evolution will allow us to check that our adopted K-corrections are
valid (a potential systematic uncertainty that affects both teams in
common, as the K-corrections are based on the same, small sample of
nearby SN spectrophotometry). Additionally, the $U$-band observations
are particularly interesting; they provide great leverage on measuring
variations in the extinction law. However, in order to make sense of
the rest-frame $U$-band data at high redshift, we also need a
comparison sample of $U$-band light curves of \emph{nearby} SN Ia.
Such a sample is becoming available only now, with $UBVRI$ light
curves of about 40 SN Ia observed as part of a monitoring campaign at
the CfA (Jha et al.~2001, in preparation).

\section{Evolution}

If the observed faintness of high-redshift SN Ia cannot be explained
by intervening material along the line of sight, intrinsic variations
in the luminosities of the supernovae themselves may be the culprit.
Without a detailed understanding of SN Ia and their progenitors,
predicting the expected evolutionary effects to $z \simeq 0.5$ is
difficult. So far there is no indication that the SN Ia observed at
high-redshift are intrinsically different from those nearby, with the
light curves and spectra showing a strong similarity (R98; P99; Coil
et al.~2000). More observations may strengthen this case; for instance
the rest-frame $U$-band data from the current \emph{HST} campaign may
be useful in this regard, as variations in SN Ia progenitors are
predicted to lead to observable effects in the near ultraviolet
(H\"oflich et al.~2000). However, it is unclear whether the data will
be precise enough to detect subtle indications of evolution. Beyond
this, we do not have clear knowledge of how potential evolution of SN
Ia would affect their luminosities (and more importantly, the
luminosity/decline-rate relationship), which are the basis for the
cosmological inferences (see Leibundgut 2000 for a recent review).

\begin{figure}
\plotone{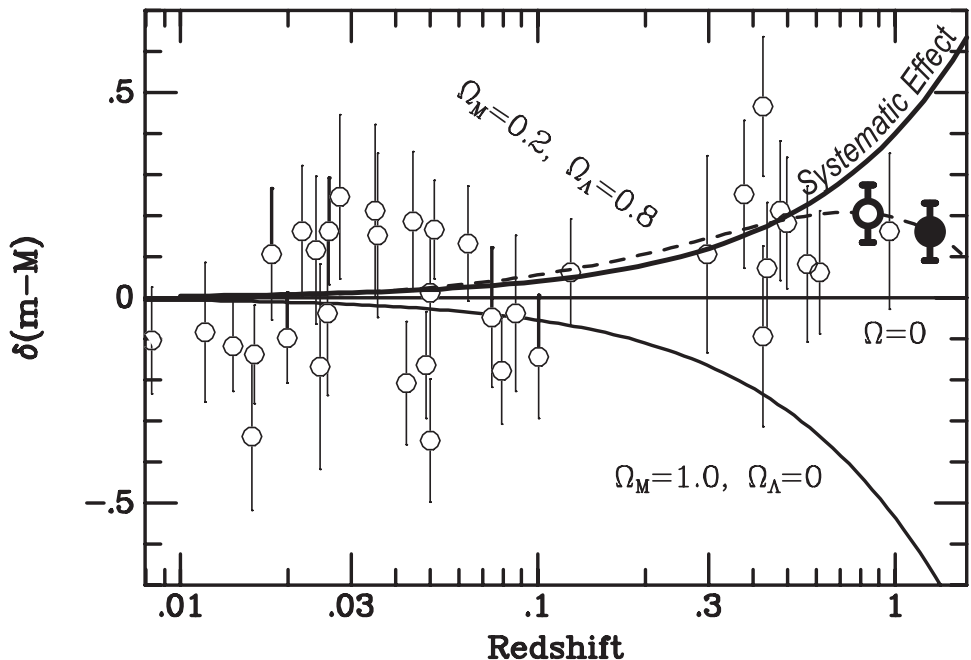}
\caption{Magnitude residuals for SN Ia relative to an empty
Universe. The expected magnitude-redshift relation from an
$\Omega_\Lambda$-dominated cosmology diverges from the effects of a
systematic uncertainty. Observations of SN Ia at higher redshift, $z
\simeq 1$ will allow us to distinguish these possibilities. The heavy
open and filled points show the expected uncertainties from samples of
SN Ia at $z \simeq 0.8$ and $z \simeq 1.2$ currently being obtained.}
\end{figure}

Because of the difficulty of definitively ruling out evolution or
other systematic uncertainties, it would be desirable to test directly
that the SN Ia are faint because of cosmology. Fortunately, the
best-fit cosmologies in Figure 1 provide an avenue for such a
test. In a flat $\Omega_M = 0.3$, $\Omega_\Lambda = 0.7$
Universe there was a transition from decelerated expansion (where the
energy density was dominated by matter) to accelerated expansion that
occurred at $z \simeq 0.7$. This effect is shown in Figure 4, where we
display the luminosity-distance/redshift relation for various
cosmologies, relative to an empty Universe. The heavy dashed line
shows the expectation for an $\Omega_M = 0.2$, $\Omega_\Lambda = 0.8$
model, and it shows that at higher redshifts $z \simeq 1$, the SN
become less faint relative to an empty Universe, i.e. the prediction
turns over. If the current SN data were explained by other effects,
for instance a systematic uncertainty which grows linearly with
redshift illustrated as by the heavy solid line, we would expect the SN
Ia results at higher redshift to diverge significantly from the
cosmological prediction. Thus, observations of SN Ia at $z \simeq 1$
provide an excellent test to confirm or refute the current
cosmological paradigm.

The downside is that finding and following up SN Ia at $z \simeq 1$
with the necessary precision is quite a bit more difficult than
measuring SN Ia at $z \simeq 0.5$. Both groups have undertaken
efforts to observe such SN Ia (e.g., Fabbro et al.~1999; Tonry et
al.~1999), and the reduction and analysis of these objects are
underway. More objects are necessary, and we in the High-Z Team
will make these higher-redshift SN Ia a priority in the upcoming
year. With these data, we expect to measure the
luminosity-distance/redshift relation at $z \simeq 0.8$ and $z \simeq
1.2$ with the uncertainties shown by the heavy open and filled points
in Figure 4. This should allow us to discriminate the cosmological
signal from systematic effects.

Other lines of astrophysical evidence suggest that we live in a flat
Universe with low matter density ($\Omega_M \simeq 0.3$), with a
dominant contribution from ``dark energy'' such as the cosmological
constant. But \emph{only} the data from SN Ia provide clear and direct
evidence for the qualtiative prediction of dark energy: acceleration
of the expansion. It is imperative, then, that the SN Ia results be
checked against systematic error, which is the purpose of the
observations discussed here. If the SN Ia results indeed arise from
some systematic uncertainty, resolving the current cosmological
constraints will be an exciting challenge. On the other hand,
confirming the SN Ia results will open up even more exciting questions
as to the nature and evolution of the mysterious dark energy. It is
quite likely that continued precise observations of SN Ia will provide
an important tool for these further investigations.

\acknowledgements

We thank Andrew Jaffe for providing the CMB likelihood distribution
used in Figure 1. We also express our appreciation to the \emph{HST}
and numerous ground-based telescope time allocation committees for
support. This research is also supported at Harvard by grant AST
98-19825 from the National Science Foundation and by grant GO:08648
provided by NASA through the Space Telescope Science Institute which
is operated by AURA under NASA contract NAS 5-26555.

\end{document}